\documentclass{aa}
\usepackage{graphics}
\input psfig.sty

\newcommand{\ltsima} {$\; \buildrel < \over \sim \;$}
\newcommand{\gtsima} {$\; \buildrel > \over \sim \;$}
\newcommand{\lta} {\lower.5ex\hbox{\ltsima}}
\newcommand{\gta} {\lower.5ex\hbox{\gtsima}}

\begin{document}

\thesaurus{03 (11.01.2; 11.05.1; 11.10.1; 11.14.1)}
\title{The HST view of FR~I radio galaxies:\\ 
evidence for non-thermal nuclear sources
\thanks{Based on observations with the NASA/ESA Hubble Space Telescope, 
obtained at the
Space Telescope Science Institute, which is operated by AURA, Inc.,
under NASA contract NAS 5-26555 and by STScI grant GO-3594.01-91A}}

\author{M. Chiaberge \inst{1}, A. Capetti \inst{2} 
\and A. Celotti \inst{1}}

\offprints{M. Chiaberge}

\institute{SISSA/ISAS, Via Beirut 2-4, I-34014 Trieste, Italy \\
email: chiab@sissa.it
\and Osservatorio Astronomico di Torino, Strada Osservatorio 20, 
I-10025 Pino Torinese, Italy}

\date{Received 27 January 1999; accepted ...}

\titlerunning{HST observations of FR~I radio galaxies}
\authorrunning{Chiaberge et al.}  \maketitle
\begin{abstract}

Unresolved nuclear sources are detected by
the Hubble Space Telescope in the great majority of a complete
sample of 33 FR~I radio galaxies belonging to the 3CR catalogue.

The optical flux of these Central Compact Cores (CCC) shows a 
striking linear correlation with the
radio core one over four decades, arguing for a non--thermal
synchrotron origin of the CCC radiation.  We also find evidence that
this emission is anisotropic, which leads us to identify 
CCCs with the misoriented relativistic jet component which dominates
in BL Lac objects.  This interpretation is also supported by the
similarity in the radio-to-optical and optical spectral indices.

The high rate of CCC detection (85 \%) 
suggests that a ``standard'' pc scale, geometrically thick torus is
not present in low luminosity  radio-galaxies.
Thus the lack of broad lines in FR~I cannot be attributed to obscuration.

CCC fluxes also represent upper limits to any thermal/disc emission.
For a $10^9 M_{\sun}$ black hole, typical of FR~I sources, these limits
translate into a fraction as small as $\lta 10^{-7}-10^{-5}$ of the
Eddington luminosity.


\end{abstract}

\section{Introduction}
\label{intro}
Optical studies of radio galaxies are central for the understanding of
the physics of their nuclei, by allowing to investigate the
relationship between the environment/host galaxy and the occurrence of
activity (e.g. the role of merging in the evolution of the nuclear
fueling, Colina \& De Juan \cite{colina}), the formation of jets, the
`schemes' unifying different classes of Active Galactic Nuclei (AGN) and in 
particular radio
galaxies with object dominated by relativistic beamed emission
(blazars) (e.g. Urry \& Padovani \cite{urry95}).

The original classification of radio galaxies by Fanaroff \& Riley
(\cite{fr}) is based on a morphological criterion, i.e. edge
darkened (FR~I) vs edge brightened (FR~II) radio structure. 
It was later discovered
that this dicothomy corresponds to a (continuous) transition in total
radio luminosity (at 178 MHz) which formally occurs at $L_{178}=
2\times 10^{26}$ W Hz$^{-1}$. In this paper we focus on the properties
of FR~I radio galaxies, which represent the lower power objects.

From the optical point of view FR~I are associated with elliptical
galaxies, are generally found in regions of galaxy density higher than
powerful radio sources (Zirbel \cite{zirb97}) and often show signs of
interactions (Gonzales-Serrano et al., \cite{gonz93}).  Their optical
spectra are dominated by starlight with no evidence for a continuum
component directly related to the active nucleus.  In general faint
narrow lines are present while no broad lines are detected (Morganti
et al. \cite{morg92}, Zirbel \& Baum \cite{zirb95}).

Significant progresses in the understanding of the inner structure of
FR~I have been obtained thanks to HST observations. Most importantly
they revealed the presence of dusty or gaseous kpc scale disks in the
nuclear regions of several FR~I radio galaxies (Ford et
al. \cite{ford94}; Jaffe et al. \cite{jaff93}; De Koff et
al. \cite{deko95}; Van der Marel, \& Van den Bosch \cite{vand98}).
The study of the dynamics of these disks provides one of the strongest
pieces of evidence to date of the presence of supermassive black holes
associated with the activity in galactic nuclei, with masses
reaching $\sim 10^9 M_{\sun}$ (Harms et al. \cite{harm94},
Macchetto et al. \cite{macc97}, Ferrarese et al. \cite{ferr96}, Van
der Marel \& Van den Bosch \cite{vand98}, Bower et al. \cite{bowe98}).

A further, newly discovered feature in FR~I, which will concentrate
on, are faint, nuclear optical components, which might represent the
(as yet) elusive emission associated with the AGN. Their study can be a
powerful tool to directly compare the nuclear properties of FR~I with
those of other AGNs, such as BL Lac objects and powerful radio
galaxies.

In fact, in the frame of the unification schemes of low luminosity
radio loud AGN, FR~I radio galaxies are believed to be the misaligned
counterparts of BL Lacs (for a review, see Urry \& Padovani 
\cite{urry95}). In this
scenario the non--thermal beamed emission from relativistic jets,
which dominates in blazars, should also be present in radio galaxies
although not amplified or even de-amplified.  The possibility of
directly detecting this component in the optical band thanks to the
HST capabilities, has been explored by Capetti \& Celotti
(\cite{ac2}).  They studied five radio galaxies whose extended nuclear
discs can be used as indicators of the radio sources orientation. The
ratio of the nuclear luminosities of FR~I and BL Lacs with similar
extended properties, shows a suggestive correlation with the
orientation of the radio galaxies.  This behavior is quantitatively
consistent with a scenario in which also the FR~I emission is
dominated by the beamed radiation from a relativistic jet.  Further,
independent support to this interpretation comes from the rapid
variability of the central source of \object{M~87}, the only object
for which multi epoch HST data were available (Tsvetanov et
al. \cite{tsve98}).

The role of obscuration in low luminosity radio galaxies is also still a 
matter of debate. 
Within the unification scheme for Seyfert galaxies, it is believed
that in type 2 objects the Broad Line Region (BLR) and the nuclear
continuum source are hidden by an absorbing, edge-on torus, while in
Seyfert 1 our line of sight is within absorption-free visibility cones
(Antonucci \& Miller \cite{anto85}). Similarly, a combination of
obscuration and beaming is essential for the unification of powerful
radio sources (Barthel \cite{bart89}).  However, although
circumnuclear tori appear to be commonly associated with active
galactic nuclei, there is as yet no evidence in favour of
nuclear obscuring material in FR~I (and this is not required by the
FR~I / BL Lac unified scheme, Urry \& Padovani \cite{urry95}).  A
search for H$_2$O megamasers (which have been successfully used to
probe the dense molecular gas associated to the torus in Seyfert
galaxies, Miyoshi et al. \cite{miyo95}, Braatz et al. \cite{braa96})
in a sample of FR~I galaxies gave negative results (Henkel et
al. \cite{henk98}).

In order to further and thoroughly investigate these issues, we
consider a complete sample of 33 3CR radio galaxies, morphologically
identified as FR~I sources. For 32 of these, HST/WFPC2 images are
available in the public archive. Most of the images were taken as part
of the HST snapshot survey of the 3C radio source counterparts, and
are already presented by Martel et al. (\cite{martel}) (objects with
$z<0.1$) and by De Koff et al. (\cite{De Koff}) (objects with
$0.1<z<0.5$). Here we specifically focus on the origin of these 
unresolved nuclear sources, which we found to be present in most of the 
objects of the sample and to which we refer as Central Compact Cores (CCC).

The selection of the sample is presented and discussed in
Sect. \ref{thesample}, while in Sect. \ref{hstobs} we describe the HST
observations.  In Sect. \ref{CCC} and Sect. \ref{origin} we focus on the
detection and origin of the CCC component, respectively and in
Sect. \ref{discussion} we discuss some of the consequences of our
results. Our findings are summarized in the final Sect. \ref{summary}.

\section{The sample}
\label{thesample}

\begin{table*} 
\caption{Summary of radio and optical data of the sample.}
\hspace{1.5cm} 
\begin{tabular}{l c c c c c c} \hline

Name    &  Other name  &   Redshift   & $F_c$ (5 GHz) &  ref. & $F_t$ (178 MHz)  &  $L_t$ (178 MHz)  \\
        &              &     $z$      &      mJy     &       &       Jy               &      W Hz$^{-1}$   \\
				
\hline
		
3C~028      &           &   0.1952     &  $<$0.2      & G88 &    16.3      & 26.94    \\
3C~029      & UGC~595   &   0.0448     &    93.0      & M96 &    15.1      & 25.71     \\    
3C~031      & NGC~383   &   0.0169     &    92.0      & G88 &    16.8      & 24.93      \\  
3C~066B     &           &   0.0215     &   182.0      & G88 &    24.6      & 25.30     \\ 
3C~075      &           &   0.0232     &    39.0      & M96 &    10.5      & 25.00     \\
3C~76.1     &           &   0.0324     &    ---       &     &    12.2      & 25.35     \\
3C~078      & NGC~1218  &   0.0288     &   964.0      & M96 &    19.75     & 25.46    \\
3C~083.1    & NGC~1265  &   0.0251     &    21.0      & R75 &    26.6      & 25.47     \\
3C~084      & NGC~1275  &   0.0176     &   42370.0    & T96 &    40.5      & 25.32    \\
3C~089      &           &   0.1386     &    49.0      & Z95 &    20.2      & 26.76       \\
3C~264      & NGC~3862  &   0.0206     &   200.0      & G88 &    26.0      & 25.29     \\
3C~270      & NGC~4261  &   0.0074     &   308.0      & G88 &    55.45     & 24.73     \\
3C~272.1    & M~84      &   0.0037     &   180.0      & G88 &    19.4      & 23.68    \\
3C~274      & M~87      &   0.0037     &  4000.0      & G88 &  1050.0      & 25.42    \\
3C~277.3    & COMA A    &   0.0857     &    12.2      & G88 &     9.0      & 26.03     \\
3C~288      &           &   0.2460     &    30.0      & G88 &    18.9      & 27.17     \\
3C~293      & UGC~8782  &   0.0452     &   100.0      & G88 &    12.7      & 25.65     \\
3C~296      & NGC~5532  &   0.0237     &    77.0      & G88 &    13.0      & 25.11     \\ 
3C~305      & IC~1065   &   0.0414     &    29.5      & G88 &    15.7      & 25.66     \\
3C~310      &           &   0.0540     &    80.0      & G88 &    55.1      & 26.43     \\
3C~314.1    &           & 0.1197       &  $<$1.0      & G88 &    10.6      & 26.37     \\
3C~315      &           & 0.1083       &   150.0      & G88 &    17.8      & 26.51     \\
3C~317      & UGC~9799  & 0.0342       &   391.0      & M96 &    47.3      & 25.98     \\
3C~338      & NGC~6166  & 0.0303       &   105.0      & G88 &    46.9      & 25.87     \\
3C~346      &           & 0.1620       &   220.0      & G88 &    10.9      & 26.62    \\ 
3C~348      & Her A     & 0.1540       &     0.01     & M96 &   350.0      & 28.09    \\ 
3C~386      &           & 0.0170       &    14.       & S78 &    23.9      & 25.09     \\
3C~424      &           & 0.1270       &    18.       & B92 &    14.0      & 26.54    \\
3C~433      &           & 0.1016       &     5.       & G88 &  56.2        & 26.96    \\
3C~438      &           & 0.2900       &    17.0      & Z95 &  46.3        & 27.68     \\
3C~442      & ARP~169   & 0.0262       &     2.0      & G88 &  16.1        & 25.29    \\
3C~449      & UGC~12064 & 0.0181       &    37.0      & G88 &  11.5        & 24.82    \\
3C~465      & NGC~7720  & 0.0301       &   270.0      & G88 &  37.8        & 25.78     \\
                				  
\hline

\end{tabular}
\label{tab_lett}

\medskip
$F_c$ and $F_t$ are the core and total radio fluxes. 
In column 5 we report the references to the core fluxes.
B92: Black et al. \cite{black92}, G88: Giovannini et al. \cite{giovannini}, 
L91: Leahy \& Perley \cite{leahy91}, M96: Morganti et al. \cite{morg93}, 
R75:  Riley \& Pooley \cite{riley}, 
T96: Taylor et al. \cite{tay96}, Z95: Zirbel \& Baum \cite{zirb95}, 
S78: Strom et al. \cite{strom}.
The radio core flux of \object{3C~424} has been estimated from the contour map 
(observations at 8.3 GHz).
\end{table*}


Our sample comprises all radio galaxies belonging to the 3CR catalogue
(Spinrad et al. \cite{spinrad}) and morphologically identified as FR~I
radio sources by Laing et al. (\cite{laing}) and/or Zirbel \& Baum
\cite{zirbel} (see Table \ref{tab_lett}).

However, the powerful but simple original morphological FR~I/II
classification has revealed to be often inadequate, as being somewhat
subjective and sensibly depending on the quality, resolution and
frequency of the available radio maps. Also, several radio sources
show a complex morphology: for example signatures of FR~I structures
(such as extended plumes and tails) can be detected together with
typical characteristics of FR~II sources (narrow jets and hot spots)
in sources which might represent transition FRI/II objects (see
e.g. Parma et al.  \cite{parm87}, Capetti et al. \cite{capetti95}).
Furthermore, even among the edge darkened radio galaxies, a large
variety of structures is present, including wide and narrow angle
tails, fat doubles and twin jet sources. Although all these objects
are classified as FR~I and share the common characteristic of a low
total radio luminosity, it is far from obvious that they represent a
well defined class.

Therefore, in order to establish possible differences among the
optical properties of the various subclasses of FR~I galaxies and also
directly re-assess their radio morphology against erroneous or
doubtful identifications, we searched the literature for recent radio
maps of each object of our sample.  The radio structure of at least
four sources is peculiar, there are several transition FR~I/II objects
and each of the FR~I morphological `types' described above is
represented in the sample.  In view of this ambiguity of the simple
morphological classification, in the following we will also consider
separately objects below and above a total radio luminosity of
$L_{178}= 2\times 10^{26}$ W Hz$^{-1}$, i.e.  the fiducial radio power
separation between FR~I and FR~II.  Two thirds of the sources of our
sample lie below this value.

Having only excluded \object{3C~231} (M~82) from the original list, as
it is in fact a well known starburst galaxy, the remaining radio
galaxies constitute a complete, flux limited sample of 33 FR~I
sources.  In Table \ref{tab_lett} we report redshifts, 
radio fluxes and total luminosities as taken from the literature:
redshifts span the range $z = 0.0037$ -- $0.29$, with
a median value of $z= 0.03$, and total radio luminosities at 178 MHz
are between $10^{23.7}$ and 10$^{28.1}$ W Hz$^{-1}$ ($H_0= 75$ km
s$^{-1}$ Mpc$^{-1}$ and $q_0=0.5$ are adopted hereafter).

\section{HST observations}
\label{hstobs}

HST observations are available in the public archive (up to July 1998)
for 32 out of the 33 sources (only \object{3C~76.1} has not been
observed).  The HST images were taken using the Wide Field and
Planetary Camera 2 (WFPC2).  The pixel size of the Planetary Camera,
in which the target is always located, is 0$^{\prime\prime}$.0455 and
the 800 $\times$ 800 pixels cover a field of view of
$36^{\prime\prime} \times 36^{\prime\prime}$.  The whole sample was
observed using the F702W filter as part of the HST snapshot survey of
3C radio galaxies (Martel et al. \cite{martel}, De Koff et al. \cite{De
Koff}). For about half of the sources, additional archival images
taken through narrow and broad filters are also available.  The HST
observations log is reported in Table \ref{campione}.

\begin{table*}
\caption{Log of HST observations}
\hspace{1.5cm} \begin{tabular}{l l c c   c c c  c  c  l l c c} \hline
Name     &  Filter   & t$_{\rm exp}$ (s) & Date   & & & & & & Name   &    Filter  & t$_{\rm exp}$ (s) & Date \\   \hline
3C~28    &  F702W    &    280    &   17/10/94   & & & & & &   3C~277.3   &  F702W  &   560   &   16/03/94  \\  
3C~29    &  F702W    &    280    &   12/01/95   & & & & & &	         &  F555W  &   600   &   06/06/96  \\       
3C~31    &  F702W    &    280    &   19/01/95   & & & & & &	         &  FR680N &   600   &   27/11/95  \\     
	 &  FR680N   &    600    &   01/09/95   & & & & & &	3C~288   &  F702W  &   280   &   30/04/95  \\      
3C~66B   &  F702W    &    280    &   18/03/94   & & & & & &	3C~293   &  F702W  &   280   &   15/01/95  \\         
3C~75    &  F547M    &    900    &   10/03/97   & & & & & &	3C~296   &  F702W  &   280   &   14/12/94  \\       
	 &  F791W    &    750    &   10/03/97   & & & & & &	3C~305   &  F702W  &    560  &   4/09/94   \\      
	 &  F673N    &   2500    &   10/03/97   & & & & & &	3C~310   &  F702W  &   280   &   12/09/94  \\       
3C~76.1  &   ---     &    ---    &    ----      & & & & & &     3C~314.1 &  F702W  &   280   &   22/12/94  \\
3C~78    &  F702W    &    280    &   17/08/94   & & & & & &	3C~315   &  F702W  &   280   &   29/11/94  \\           
         &  F673N    &    600    &   07/08/95   & & & & & &	3C~317   &  F814W  &  6500   &   10/08/97  \\          
	 &  F555W    &    600    &   16/09/96   & & & & & &	         &  F555W  &  6200   &   10/08/97  \\        
3C~83.1  &  F702W    &    280    &  22/07/94    & & & & & &	3C~338   &  F702W  &   280   &   09/09/94  \\         
         &  F673N    &    600    &  22/10/95    & & & & & &	         &  F673N  &   600   &   27/11/95  \\         
3C~84    &  F702W    &    560    &  31/03/94    & & & & & &	3C~346   &  F702W  &   280   &   01/08/94  \\       
         &  F450W    &    200    &  16/11/95    & & & & & &	3C~348   &  F702W  &   280   &   09/05/94  \\           
3C~89    &  F702W    &    280    &  02/07/94    & & & & & &	3C~386   &  F702W  &   280   &   1/04/ 95  \\       
3C~264   &  F547M    &   900     &  19/05/96    & & & & & &	3C~424   &  F702W  &   280   &   05/09/94   \\           
         &  F673N    &   2500    &  19/05/96    & & & & & &	3C~433   &  F702W  &   280   &   26/04/95  \\       
         &  F791W    &   750     &  19/05/96    & & & & & &	3C~438   &  F702W  &   300   &   15/12/94  \\      
3C~270   &  F791W    &   800     &  13/12/94    & & & & & &	3C~442   &  F791W  &   280   &   29/12/94  \\      
         &  F547M    &   800     &  13/12/94    & & & & & &	         &  F673N  &   2500  &   20/04/96  \\      
3C~272.1 &  F814W    &   520     &  04/03/96    & & & & & &	         &  F547M  &   900   &   20/04/96  \\         
         &  F547M    &   1200    &  04/03/96    & & & & & &	3C~449   &  F702W  &   280   &   06/08/94  \\     
         &  F658N    &   2600    &  04/03/96    & & & & & &	         &  FR680N &   600   &   24/08/95  \\         
3C~274   &  F814W    &   30      &  03/02/95    & & & & & &	3C~465   &  F702W  &   280   &   23/01/95  \\        
         &  F547M    &   30      &  03/02/95    & & & & & &              &  F673N  &   600   &   23/08/95  \\
\hline

\hline
\end{tabular}
\label{campione}
\end{table*}

The data have been processed through the PODPS (Post Observation Data
Processing System) pipeline for bias removal and flat fielding
(Biretta et al. \cite{biretta}). Individual exposures in each filter
were combined to remove cosmic rays events.

\begin{figure*}
\centerline{\psfig{figure=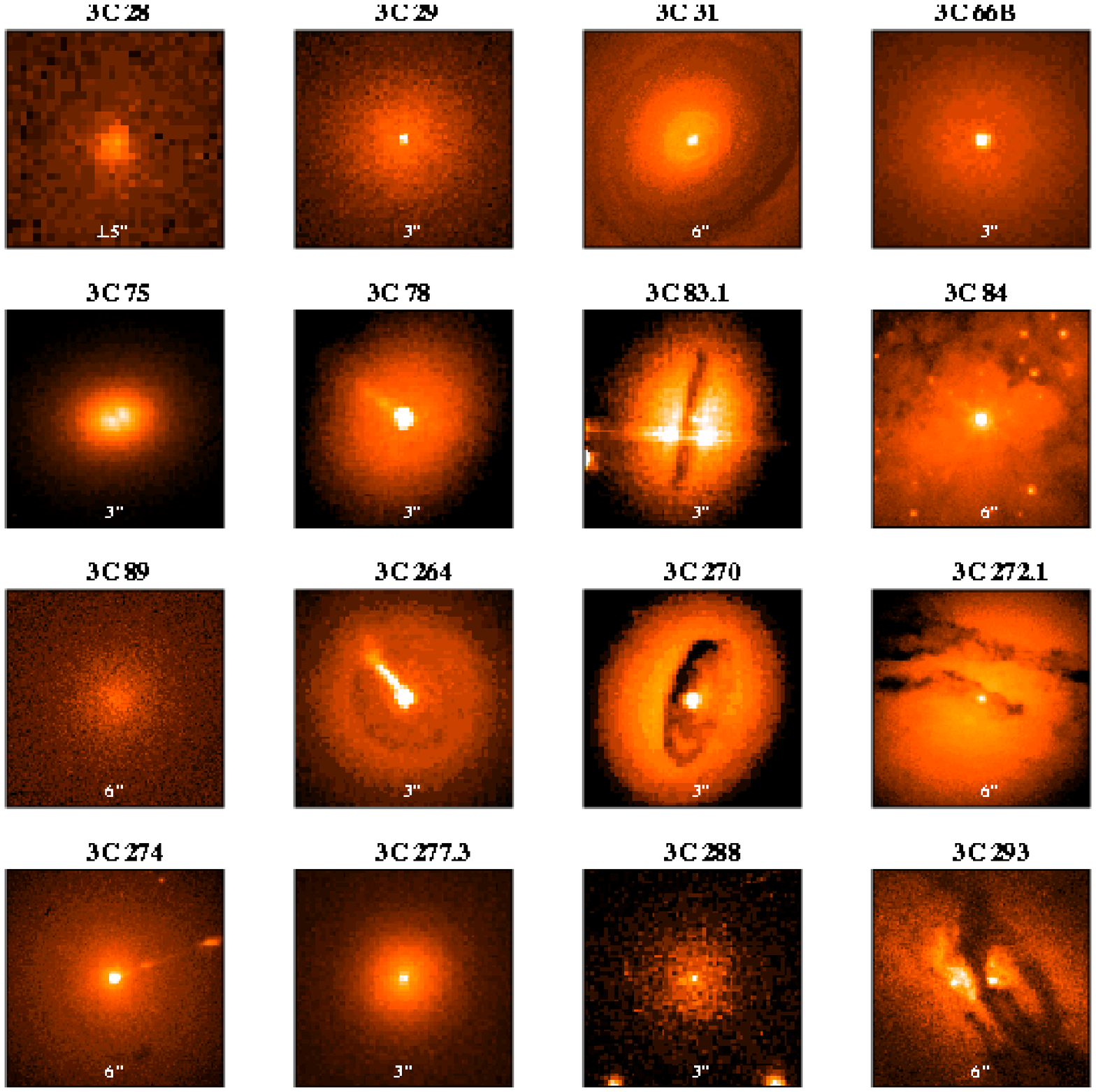,width=0.95\textwidth,angle=0}}
\caption{HST/WFPC2 broad band images of the FR~I radio galaxies of the
sample (a).}
\label{figurine1}
\end{figure*}

\begin{figure*}
\centerline{\psfig{figure=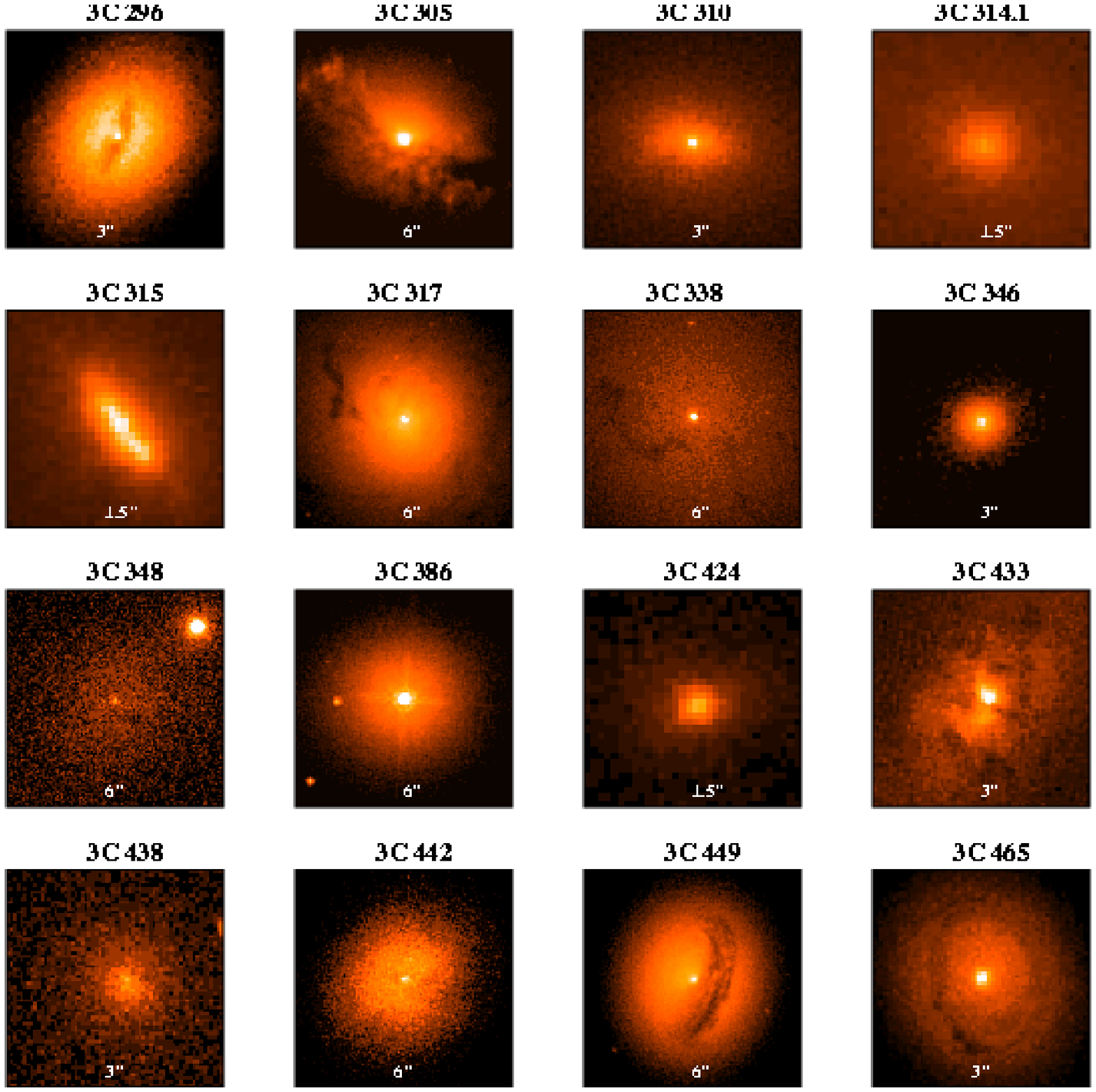,width=0.95\textwidth,angle=0}}
\caption{HST/WFPC2 broad band images of the FR~I radio galaxies of the
sample (b).}
\label{figurine2}
\end{figure*}

In Figs. \ref{figurine1}, \ref{figurine2} we present the final broad
band images of the innermost regions (1.5$^{\prime\prime}$ -- 6
$^{\prime\prime}$) of our 32 radio galaxies.  The most interesting
feature which is present in the great majority of them, is indeed an 
unresolved central source.

\section{The Central Compact Cores} 
\label{CCC}

\subsection{Identification of CCCs}
\label{detection}

While it is straightforward to identify unresolved sources when they
are isolated, the situation is more complex when these are located at
the center of a galaxy, superposed to a brightness distribution with
large gradients and whose behaviour in the innermost regions cannot be
extrapolated from its large scale structure with any degree of
confidence.

We therefore adopted a simple operative approach: we derived the
radial brightness profiles of the nuclear regions of all galaxies
using the IRAF {\it RADPROF} task and measured the FWHM setting the
background level at the intensity measured at a distance of $\sim$5 pixels
($\sim$0.23$^{\prime\prime}$) from the center.  In 22 cases the measured
FWHM is in the range 0.05$^{\prime\prime}$ -- 0.08$^{\prime\prime}$,
i.e. indicative of the presence of an unresolved source at the HST
resolution.

\begin{table}
\caption{Summary of the CCC photometry}
\begin{tabular}{l r r} \hline

Name    &        CCC flux, filter 1       &          CCC flux, filter 2                \\
        & erg s$^{-1}$cm$^{-2}$\AA$^{-1}$ &   erg s$^{-1}$ cm$^{-2}$ \AA$^{-1}$      \\
\hline										    
										    
3C~028   &   $<$3.2 E-19 $^a$              &                                     \\
3C~029   &   $(5.8  \pm 0.7)$  E-18 $^a$   &	 				 \\
3C~031   &   $(1.5  \pm 0.4)$  E-17 $^b$   &                                     \\
3C~066B  &   $(4.93 \pm 0.07)$ E-17 $^a$   &	 				 \\
3C~075   &        ------~~~~~~~~~  	   &					 \\
3C~078   &   $(2.38 \pm 0.04)$ E-16 $^b$   &  $(3.09 \pm 0.05)$ E-16 $^e$        \\
3C~083.1 &   $1.4$ E-18:                   &                                     \\
3C~084   &   $(1.5  \pm 0.3)$  E-15 $^b$   &  $(1.43 \pm 0.07)$ E-15 $^g$        \\
3C~089   &       $<2$ E-19                 &	  				 \\
3C~264   &   $(1.14 \pm 0.06)$ E-16 $^c$   &  $(1.54 \pm 0.05)$ E-16 $^f$        \\
3C~270   &   $(5.1 \pm 0.1)$   E-18 $^c$   &  $(6.0 \pm 2.0)$   E-18 $^f$ 	 \\
3C~272.1 &   $(5.9 \pm 0.2)$   E-17 $^d$   &  $(4.2 \pm 0.1)$   E-17 $^f$        \\
3C~274   &   $(3.9 \pm 0.2)$   E-16 $^d$   &  $(6.5 \pm 0.2)$   E-16 $^f$        \\
3C~277.3 &    $(1.5 \pm 1.2)$  E-18 $^b$   &  $(3.5 \pm 1.2)$   E-18 $^e$        \\
3C~288   &    $(7.0 \pm 1.0)$  E-19 $^a$   &		 	 		 \\
3C~293   &       ------~~~~~~~~~~  	   &					 \\
3C~296   &    $(3.4 \pm 0.3)$ E-18 $^a$	   &					 \\
3C~305   &           ------~~~~~~~~~~      &					 \\
3C~310   &    $(3.5 \pm 0.7)$ E-18 $^a$    &					 \\
3C~314.1 &    $<$9.5  E-19 $^a$            &					 \\
3C~315   &       ------~~~~~~~~~~  	   &					 \\
3C~317   &    $(9.6 \pm 1.0)$ E-18 $^d$    &  $(1.5 \pm 0.2)$   E-17 $^e$	 \\
3C~338   &    $(1.0 \pm 0.1)$ E-17 $^b$    &                                     \\
3C~346   &    $(2.3 \pm 0.3)$ E-17 $^a$	   &					 \\
3C~348   &    $(8.0 \pm 1.0)$ E-19 $^a$	   &					 \\
3C~386   &    $(1.4 \pm 0.1)$ E-15 $^a$	   &					 \\
3C~424   &    $<$1.5 E-18 $^a$             &					 \\
3C~433   &        ------~~~~~~~~~~ 	   &					 \\
3C~438   &     $<$4.0 E-19 $^a$            &			                 \\
3C~442   &    $(9.1 \pm 3.3)$ E-19 $^a$    &  $(8.0 \pm 2.0)$  E-19 $^f$         \\
3C~449   &    $(1.8 \pm 0.1)$ E-17 $^b$    &                                     \\
3C~465   &    $(1.9 \pm 0.1)$ E-17 $^b$    &                                     \\

\hline

\end{tabular}
\label{tab_ccc}

\medskip

HST filters coding: 
$^a$ F702W, $^b$ F702W line subtr., $^c$ F791W, $^d$ F814W, $^e$ F555W, $^f$ F547M, $^g$ F450W

\end{table}

In 5 cases, namely \object{3C~28}, \object{3C~89}, \object{3C~314.1},
\object{3C~424} and \object{3C~438}, the fitting procedure yields FWHM
larger than $0.15^{\prime\prime}$.  The behaviour of these sources is
radically different from those in which a CCC is detected and
therefore we believe that, even with this operative definition, no
ambiguity exists on whether or not a central unresolved source is
present.

The remaining 5 sources have complex nuclear morphologies. The central
regions of \object{3C~75}, \object{3C~293}, \object{3C~305} and
\object{3C~433} are covered by dust lanes.  Bright compact knots are
seen, but they are completely resolved and offset from the center of
the galaxy.  Conversely \object{3C~315} has a peculiar highly
elongated structure, contrasting with the typical roundness of FR~I
host galaxies, and no central source is seen. 

\subsection{CCC photometry}

The F702W transmission curve covers the wavelength range 5900 - 8200
\AA ~and thus, within our redshift range, includes the H$\alpha$ and
[N II] emission lines. To estimate the continuum emission of the CCC
we therefore preferred, when possible, to use images obtained with the
F814W or F791W filters which are relatively line-free spectral regions
up to a redshift of 0.075.

We performed aperture photometry of the 22 CCCs, adopting the internal
WFPC2 flux calibration, which is accurate to better than 5 per cent.
However, the dominant photometric error is the determination of the
background in regions of varying absorption and steep brightness
gradients, especially for the faintest CCCs, resulting in a typical
error of 10\% to 20\%.  Narrow-band images were used, where available,
to remove the line contamination from the F702W images.  We found that
the line contribution is typically 5 to 40 per cent of the total
flux measured in the F702W filter, therefore even the uncorrected
fluxes are (for our purposes) reliable estimates of the continuum
level. Optical fluxes of the CCCs are given in Table \ref{tab_ccc}.

HST images in two broad filters are available for 9 galaxies.
However, only in 5 cases (\object{3C~78}, \object{3C~84},
\object{3C~264}, \object{3C~272.1}, \object{3C~274}) the accuracy of
the photometry is sufficient to deduce reliable estimates of the
optical slope. For the last three objects data were taken
simultaneously in the two bands, which avoids uncertainties due to
possible variability.  Galactic extinction is significant (and we
corrected for it) only in the case of \object{3C~84}, for which $A_B =
0.7$.  In \object{3C~272.1} we estimated, by comparing the F547M and
the F814W images, that the extended nuclear dust lane produces an
absorption of 1 -- 2 mag in the V band.  The derived spectral indices
(corrected for reddening) are in the range $\alpha_o = 0.7-1.3$ 
($F_{\nu}\propto \nu^{-\alpha}$).

For the galaxies only showing diffuse emission we set as upper limits
the light excess of the central 3x3 pixels with respect to the
surrounding galaxy background.  For the complex sources no photometry
was performed.

\section{Origin of the Central Compact Cores} 
\label{origin}

\begin{figure} 
\resizebox{\hsize}{!}{\includegraphics{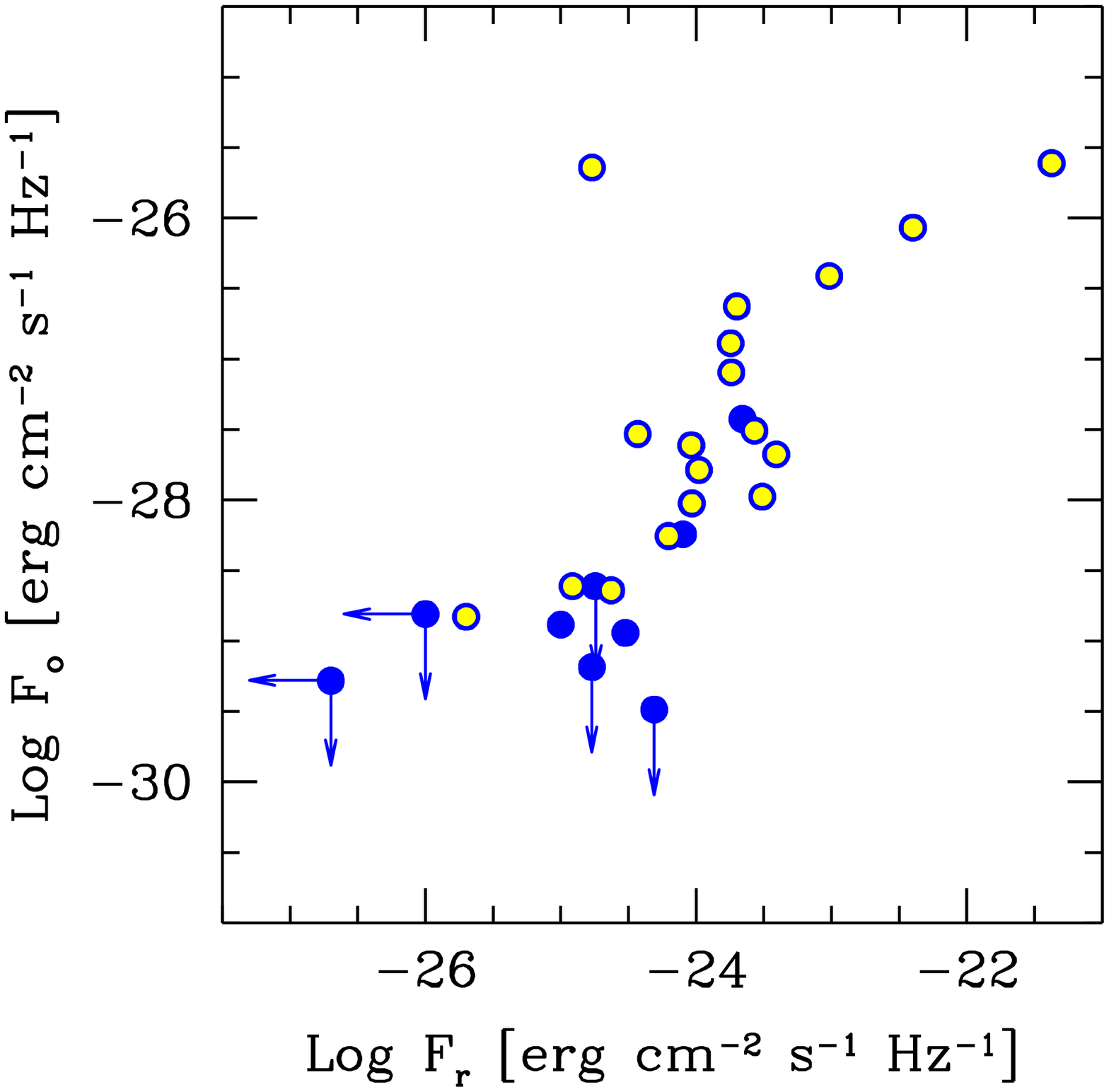}}
\caption{Optical flux of the CCC versus core radio (5 GHz) flux. Note
the clear trend between the two quantities. Only the peculiar object
3C~386 is significantly offset (see text for discussion). Again,
different symbols mark sources with total radio luminosity below (open
circles) and above (filled circles) $L_{178} = 2\times 10^{26}$ [W
Hz$^{-1}$].}
\label{flu} \end{figure}

As already pointed out in the introduction, although the presence of
nuclear optical emission associated with the active nuclei of FR~I
radio galaxies is not a surprising feature, it is important to
investigate its origin. With this aim we now explore possible
(cor-)relations between the CCC flux/luminosity and other observed
properties.

No trend is found between CCC luminosities and total radio
power or absolute visual magnitude of the host galaxy.  Conversely,
the CCCs emission bears a clear connection with the radio core. In
Fig. \ref{flu} we plot the optical flux of the CCCs $F_o$ versus the core 
radio flux $F_r$ (at $5$ GHz): a clear trend is visible.

In order to quantify this relation let us consider separately the low
and high luminosity sub--samples.  

\begin{figure}
\resizebox{\hsize}{!}{\includegraphics{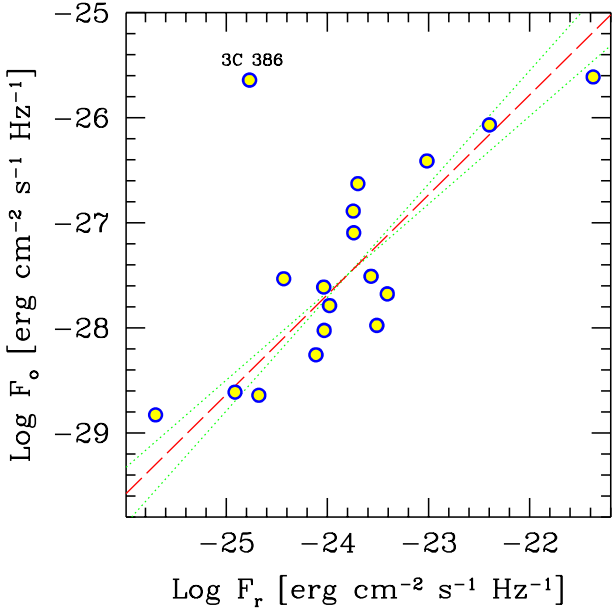}}
\caption{Optical CCC vs radio core fluxes at $5$ GHz 
for the low power subsample (see text).
Dotted lines are the fits to the data using each of the two fluxes as
independent variables. Dashed lines represent the best fits. The
peculiar object 3C~386 is excluded from the fitting procedure.}
\label{fits} 
\end{figure}

\begin{figure} 
\resizebox{\hsize}{!}{\includegraphics{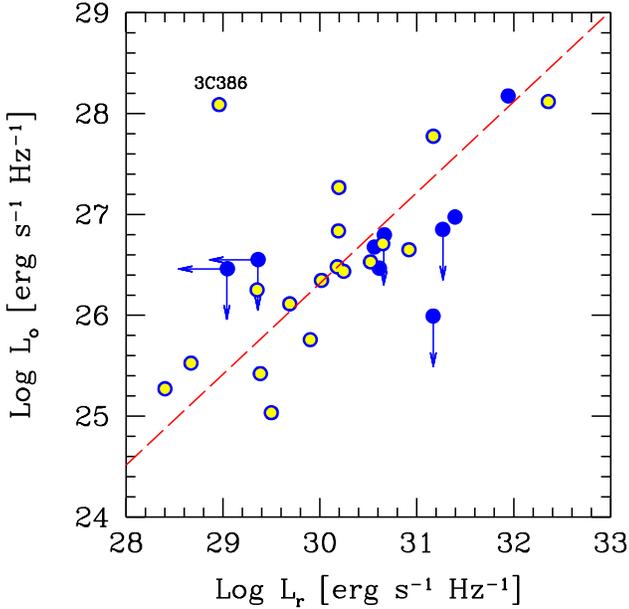}}
\caption{CCC luminosity versus radio (5 GHz) core luminosity for both
the lower (open circles) and higher (filled circles) total radio
luminosity. The dashed line is the best fit to the data of the
low luminosity subsample, having excluded the peculiar object \object{3C~386}.}
\label{lum} 
\end{figure}

The 18 CCCs associated to the low luminosity sources show a tight 
correlation between $F_o$
and $F_r$. Only one point, representing \object{3C~386}, is well
separated from the others.  We perform a non-weighted least squares fit
(excluding 3C~386): the correlation coefficient is $r=0.88$ which
gives a probability that the points are taken from a random
distribution of $P=3.1 \times 10^{-6}$. The dotted lines in
Fig. \ref{fits} are the fits to the data using each of
the two fluxes as independent variables. The best fit is represented
by the bisectrix of these two regressions (dashed line) and has a slope
of $0.95\pm 0.10$. The
statistical parameters\footnote{We also performed the linear fit by 
using a weighted Chi-square method adopting ten per cent errors on both 
variables and
a ``robust'' least absolute deviation method. 
The results both in term of probability and linear fits parameters are 
fully consistent (with even smaller errors) with those reported above.}
of the fits are reported in Table \ref{corr}. 
A similarly strong correlation ($P=6.0\times 10^{-6}$) is present also
between radio core and CCC luminosities (Fig. \ref{lum}).
The fact that the correlation is found both in flux and luminosity
gives us confidence that it is not induced by either selection effects
or a common redshift dependence. Moreover the 3CR sample has been
selected according to the total radio flux at low frequency which is
only weakly correlated with the core properties entering in the
correlations (e.g. Giovannini et al. \cite{giovannini}).

Let us then consider the case of 3C~386, which we excluded from the
statistical analysis. While all points are closely clustered around
the linear fit, with a dispersion of only $\sim$ 0.4 dex, 3C~386
deviates by 3 orders of magnitude, clearly indicating that its
optical/radio properties differ from the remaining of the sample. This
suggestion is strengthened by the presence of a further peculiarity in
the optical band of this object, i.e. the detection of a broad
H$\alpha$ line (Simpson et al. \cite{simpson}), atypical of
FR~I radio galaxies.
In any case, although the inclusion of 3C~386 in the analysis
decreases the statistical significance of the correlation to
$P=2.6\times 10^{-3}$, it would not affect our conclusions.

Turning now to the high luminosity sub-sample, five out of the nine
points are upper limits, since no CCC is detected. This prevents us
from performing a meaningful statistical analysis. Note, however, that
the four detected CCC fluxes lie along the same correlation defined by
the lower luminosity objects (see Fig. \ref{lum}).

We conclude that a striking linear correlation is present between the
core radio and CCC fluxes (and luminosities): it extends over 
four orders of magnitude, has an extremely high statistical
significance, a small dispersion and a slope consistent with unity.
Since the radio core emission is certainly originated as synchrotron
radiation the above tight link is a strong suggestion that also the
optical CCC emission is produced by the same non--thermal process.

Independent support to this hypothesis comes from considering
the spectral information relative to the CCCs, which can be compared
with that of sources where synchrotron emission dominates in the
optical as well as in the radio band (e.g. blazars, optical jets).
First of all, we find that the radio--optical spectral indices of the CCC of 
our radio galaxies span the range $\alpha_{ro} \sim 0.6$--$0.9$, 
similar to those of optical jets 
(e.g. Sparks et al. \cite{smb94}) and 
at the upper end of the spectral indeces of blazars for which 
$\alpha_{ro} \sim 0.2$--$0.9$ (e.g. Fossati et al. \cite{gfos}). 

Furthermore, the CCC optical spectral indices (determined however for
only five sources) are in the range $\alpha_o \sim 0.7-1.3$, also typical
of the synchrotron emitting sources mentioned above.

\begin{table}
\caption{Linear fit parameters}
\begin{tabular}{l c c c } 
\hline
r                         & $0.88$               & &  \\
P                         & $3.1 \times 10^{-6}$ & &  \\
$F_o$ vs $F_r$            & $a=0.84 \pm 0.12$    & $b =-7.5 \pm 2.8 $  & rms = 0.42   \\
$F_r$ vs $F_o$            & $a=0.92 \pm 0.13$    & $b = 1.7 \pm 3.5 $  & rms = 0.44   \\

\hline
\end{tabular}

\medskip

Statistical parameters for the correlation between radio core
and optical CCC fluxes.  The fitted linear relation is of the
form $y=ax+b$.
\label{corr}
\end{table}

\subsection{Alternative explanations}
\label{alternative}

Even though synchrotron radiation provides a rather convincing
interpretation of the nature of CCCs, in the following we consider and
discuss other emission processes usually occurring in the nuclei of
active galaxies. We should however already note that none of these
mechanisms seems to naturally account for the linear correlation
between the radio and optical CCC luminosities and even less
for its small dispersion, as radio cores are strongly
affected by relativistic beaming while in all the alternative
scenarios the optical emission is essentially quasi--isotropic.

$\bullet$ Nuclear cusps or star clusters: \\ 
recent works have shown that stellar concentrations are often present
in the nuclear regions of elliptical galaxies (e.g. Lauer et
al. \cite{laue95}).  However, the CCCs optical spectral slopes are not
compatible with the `red' colors typical of old stellar populations.

$\bullet$ Nuclear starburst: \\ 
as CCCs are observed in the great majority of the galaxies of our
sample, star formation should be maintained continuously for a
timescale comparable to the lifetime of the radio sources, i.e.  $\sim
10^{7-8}$ yr (Parma et al. \cite{parm99}). Although some ad hoc
mechanisms regulating the star formation rate might exist, this
possibility appears implausible.

$\bullet$ Accretion discs: \\ the optical spectral information also
contrasts with the emission expected at least in the simplest
hypothesis of a Shakura-Sunyaev geometrically thin, optically thick
disk, which generally predicts a harder spectral slope ($\alpha \lta -0.3$).
A softer index is expected only at higher frequencies (UV--soft--X;
e.g. Szuszkiewicz et al. \cite{ewa}). No observational 
constraints are provided by
more complex disc models (e.g. the recently proposed
case of low density radiatively inefficient accretion flows, ADAF,
Rees et al. \cite{rees82}; Narayan \& Yi \cite{narayan}; Chen et al. 
\cite{chen}) as they can
reproduce widely different spectral slopes in the optical band.

$\bullet$ Emission line region: \\ 
as already pointed out, although line emission contributes from 5 to 40
per cent, continuum emission provides the bulk of the total CCC flux.

We conclude from this analysis that there is no other likely
explanation for the origin of CCCs, except non--thermal synchrotron
emission.

\section{Discussion}
\label{discussion}

\subsection{Implications for unified models}
\label{unif}

For the first time, thanks to the high spatial resolution of HST, it
has been possible to separate the contribution of the host
galaxy from the genuine nuclear emission in FR~I radio galaxies, which
manifests itself as a Central Compact Core.

\begin{figure} 
\resizebox{\hsize}{!}{\includegraphics{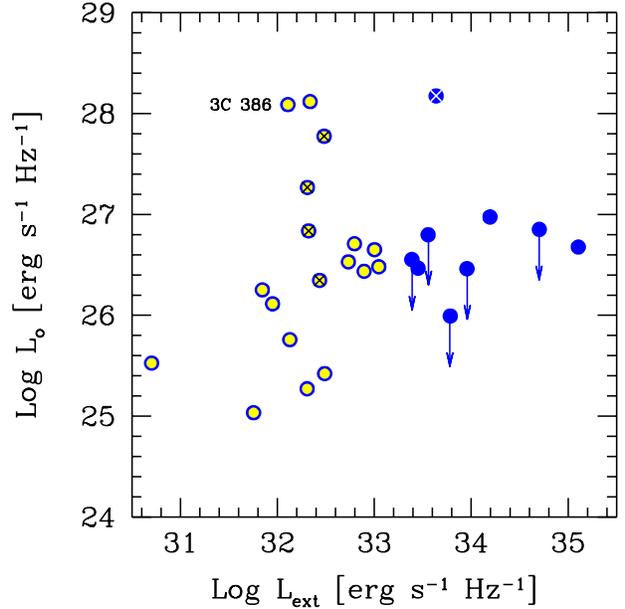}}
\caption{Optical CCC luminosity versus total radio luminosity.  Again,
different symbols mark sources with total radio luminosity below (open
circles) and above (filled circles) $L_{178} = 2\times 10^{26}$ [W
Hz$^{-1}$]. Crossed symbols are for galaxies with known optical jets.}
\label{lum_ext} 
\end{figure}

CCCs appear to be associated with non--thermal
synchrotron emission. Three pieces of evidence also suggest
that the CCC radiation is anisotropic due to relativistic beaming:

\begin{enumerate}
\item
Capetti \& Celotti (\cite{ac2}) studied five radio galaxies in which
HST images revealed the presence of extended nuclear discs, which
appear to be useful indicators of the radio sources orientation. The
FR~I CCC luminosity shows a suggestive correlation with the
orientation of the radio galaxies with respect to the line of sight;

\item
Sparks et al. (\cite{spar95}) argued that jets are detected in the
optical band only when pointing towards the observer. This would
explain why jets with optical counterparts are smaller (they are
foreshortened), brighter and one-sided (they are relativistically
beamed) with respect to typical radio jets. Five sources of our sample
(namely \object{3C~66B}, \object{3C~78}, \object{3C~264},
\object{3C~274} and \object{3C~346}) indeed show optical jets and these very
same objects clearly stand out for being among the
brightest CCC of their respective subsamples (see Fig. \ref{lum_ext} 
where we report
the CCC luminosity versus the total radio power). Sources with optical
jets are also the only FR~I galaxies detected in UV HST observations
(Zirbel \& Baum \cite{zirb98});

\item 
as noted above, the radio core emission is certainly beamed and the
corresponding orientation dependence is reflected in the large spread
found when comparing the core to extended (quasi--isotropic) radio
luminosities (Giovannini et al.  \cite{giovannini}). Therefore if the
optical emission were isotropic the $F_o$ vs $F_r$ correlation would
show at least a similarly large scatter.

\end{enumerate}

The beamed synchrotron scenario which emerges for the origin of the
CCC is therefore a strong evidence in favour of the idea that FR~I are
the misoriented counterparts of BL Lacs. The large spread (3 orders of
magnitude) in CCC luminosity among objects of similar extended
properties (see Fig. \ref{lum_ext}) can be ascribed to different
orientations.  An extensive and quantitative analysis of this issue
and of the relationship between FR~I and BL Lacs in the light of these
results will be presented in a forthcoming paper (Chiaberge et al., in
prep).

\subsection{Are there obscuring tori in FR~I?}
\label{tori}

The observation of optical synchrotron emission has important
consequences on the role/geometry of absorption structures in the
nuclear regions of such objects.  A very important result of our
analysis is in fact the high fraction of galaxies in which the central
source has been detected.

Limiting ourselves, for the moment, to the low luminosity subsample,
we found a CCC in 85 \% of the objects.  The sources in which we do
not detect it (namely \object{3C~75}, \object{3C~293} and
\object{3C~305}) are three out of the four cases in which the center
of the galaxy is affected by obscuration from a large scale dust
structure (the fourth galaxy is \object{3C~272.1} whose CCC, although
reddened, shines through the dust lane).  In order to estimate the
optical depth of obscuring material in these sources, we derived their
expected optical flux from the CCC vs radio core flux correlation. We
found that an extinction of only a few magnitudes ($A_V \lta 6$ mag,
corresponding to a column density of $N_H \lta$ 1.2 $10^{22}$ cm$^{-2}$)
is sufficient to hide the CCC optical emission.  Therefore, even in
these three cases the presence of an optically (Thomson) thick
structure is not required (although cannot be ruled out) and the CCC
emission might be simply obscured by the foreground dust. Infrared
observations can address this issue.

The behaviour of the 11 sources with higher total radio luminosity is
probably different and certainly more complex:

\noindent
-- four of them have CCC fluxes and luminosities well consistent with
the correlations found for lower power FR~I, and would simply lie on
the high luminosity part of the radio--optical correlation (see
Fig. \ref{lum})

\noindent
-- two of them show complex (possibly absorbed) nuclei;

\noindent
-- in the last five objects no unresolved component is detected (upper
limits in Figs. \ref{flu}, \ref{lum_ext}, \ref{lum}; see also
\S\ref{CCC}). The upper limits derived could be in general agreement with
the $F_o$ vs $F_r$ CCC correlation, except possibly in the case of
\object{3C~89} (the lowest point in Fig. \ref{flu}).

The smaller number of CCC found could be simply due to the fact that
this subsample is on average at higher redshift. Clearly, we cannot
exclude the alternative possibility, i.e. that these sources are
indeed obscured, which might indicate that the degree of obscuration 
increases with
the source power (a crucial test here is the inclusion in the sample
of high luminosity FR~II radio galaxies; Chiaberge et al., in prep).

The detection of CCCs indicates that we have a direct view of the
innermost nuclear regions of FR~I.  Limits on the extension of CCC
are implied by the variability of the nucleus of \object{M~87} on time
scales of two months (Tsvetanov et al. \cite{tsve98}).  Furthermore,
since the optical emission in relativistic jets is likely to be
produced co-spatially or even closer to the black hole with respect to
the radio one, we can infer a further constraint on the CCC extension
from VLBI observations. These in fact show that most of the radio
emission comes from a source unresolved at mas resolution which can be
as small as 0.01 pc, i.e.  only $\sim$ 100 Schwarzschild radii for a
$\sim$ 10$^9$ M$_{\sun}$ black hole (e.g. Junor \& Biretta
\cite{juno95}).

It therefore appears that a ``standard'', pc scale, geometrically 
thick torus with is not present in low luminosity radio
galaxies.  Any absorbing material must be
distributed in a geometrically thin structure (our CCC detection rate
implies a thickness over size ratio $\lta 0.15$) or thick tori
are present only in a minority of FR~I. This result is
particularly intriguing since dusty nuclear discs on kpc scales have
been discovered in several FR~I and they indeed are geometrically thin
(Jaffe et al. \cite{jaff96}). In this sense, the lack of broad
emission lines in FR~I cannot be accounted for by obscuration.

\subsection{Limits on thermal disc emission and AGN efficiency}
\label{disc}

Except for blazars, whose overall spectral energy distributions are
almost always dominated by the non--thermal emission from a
relativistic jet, the nuclear emission of AGNs from the optical to the
soft-X band is generally interpreted as thermal emission from
accreting material.  Conversely, we find that in FR~I a non--thermal
synchrotron component dominates the emission in all sources.  In fact,
an isotropic optical component sufficiently bright would produce a flattening
in the radio/optical CCC correlation, due to the presence of an
additional source of optical flux (with no radio counterpart), which we
do not observe (see Figs. \ref{fits} and \ref{lum}). This is surprising if
one considers that, in a
complete sample, a substantial fraction of the objects are observed at
large angles from the line of sight and thus the emission from their
jets is expected to be strongly de-beamed, favouring the detection of
any isotropic (disc) emission.

The CCC fluxes thus set upper limits to the disc emission,
which in turn imply extremely low radiative efficiency of the accretion
process.
Note, in fact, that the observed CCC emission corresponds to $\lta
10^{-7}$--$10^{-5}$ of the Eddington luminosity of a $10^{9} M_{\sun}$
black hole, which appears to be typical for these radio galaxies.
While these values argue against the presence of a radiatively
efficient accretion phase, they are still compatible with the expected
radiative cooling rate of low density and high temperature accreting
plasma in which the electron--ion coupling is ineffective and 
most of the thermal energy is thus advected inwards and not radiated (Rees
et al. \cite{rees82}; also ADAF, e.g. Narayan \& Yi
\cite{narayan}, Chen et al. \cite{chen} and ADIOS, Blandford \& Begelman 
\cite{blandford}). This latter possibility has
been indeed proposed to account for the paucity of emission in radio
galaxies harbouring supermassive black holes (Fabian \& Rees
1995). The HST observations of CCC set consistent but independent
constraints relative to the optical emission in these systems.

This low efficiency in producing thermal emission might also account
for the lack of broad lines in FR~I spectra, which could be attributed
just to the lack of those ionizing photons which, in the other classes
of active nuclei, illuminate the dense clouds forming the Broad Line
Region. And indeed Zirbel et al. (\cite{zirb95}) have
suggested the possibility that FR~I sources produce far less UV
radiation than FR~II on the basis of the comparison between emission
line and radio luminosities of radio galaxies.  Intriguingly, the only
object in which a broad line has been detected is \object{3C~386},
whose CCC presents a much larger optical luminosity with respect to
the sources with similar radio core power, possibly indicative of a
thermal contribution.

\section{Summary and conclusions}
\label{summary}

HST images of a complete sample of 33 FR~I radio galaxies belonging to
the 3CR catalogue have revealed that an unresolved nuclear source
(Central Compact Core, CCC) is present in the great majority of these
objects. 

The CCC emission is found to be strongly connected with the radio core
emission  and anisotropic. We propose that the CCC emission can be 
identified with optical
synchrotron radiation produced in the inner regions of a relativistic
jet.  Support to this possibility comes also from spectral
information. 
These results are qualitatively consistent with the unifying model in
which FR~I radio galaxies are misoriented BL Lacs objects. 

The identification of the CCC radiation with misoriented BL Lacs
emission opens the possibility of studying this class of AGNs from a
different line of sight. This can be particularly useful in
understanding the jet structure and the level of the activity occurring
near the central object whose emission in blazars is swamped by the
highly beamed component.
Further information on the nature of CCC can be inferred by
simultaneous studies of radio cores and CCC optical variability 
which could establish whether their emission is indeed
produced in the same region.  

The detection of CCC indicates that we have a direct view of
the innermost regions of the AGN ($\lta 100 R_S$).  If we restrict the
analysis to objects with a total radio power of $< 2\times 10^{26}$ 
W Hz$^{-1}$, a CCC is found in all galaxies
except three, where absorption from extended dust structures clearly
plays a role.  This casts serious doubts on the presence of obscuring
thick tori in FR~I as a whole.

Given the dominance of non-thermal emission, the CCC luminosity
represents a firm upper limit to any thermal component, which
translates into an optical luminosity of only $\lta 10^{-5}-10^{-7}$ times
the Eddington one (for a $10^{9} M_{\sun}$ black hole). This
limit on the radiative output of accreting matter is independent but
consistent with those inferred in X--rays for large elliptical
galaxies, thus suggesting that accretion might take place in a low
efficiency radiative regime (Fabian \& Rees 1995).

The picture which emerges is that the innermost structure of FR~I radio
galaxies differs in many crucial aspects from that of the other
classes of AGN; they lack the substantial BLR, tori and thermal
disc emission, which are usually associated with active
nuclei.  Similar studies of higher
luminosity radio galaxies will be clearly crucial to determine if
either a continuity between low an high luminosity sources exists or,
alternatively, they represent substantially different manifestations
of the accretion process onto a supermassive black hole.

\begin{acknowledgements}
The authors thank G. Bodo and E. Trussoni for useful comments on the manuscript
and acknowledge the Italian MURST for financial support.

This research has made use of the NASA/IPAC Extragalactic Database (NED)
which is operated by the Jet Propulsion Laboratory, California Institute of
Technology, under contract with the National Aeronautics and Space
Administration. 

\end{acknowledgements}

\end{document}